\documentclass[aps,prb,twocolumn,superscriptaddress,showpacs,floatfix]{revtex4}

\usepackage{graphicx,color}
\usepackage{dcolumn}
\usepackage{bm}
\usepackage{stmaryrd}
\usepackage{latexsym}
\usepackage{amssymb}
\usepackage{amsfonts}
\usepackage{amsmath}

\begin{document}

\title{Exotic Fractional Topological States in Two-Dimensional Organometallic Material}

\author{Wei Li}
\affiliation{Shanghai Center for Superconductivity and State Key Laboratory of Functional Materials for Informatics, Shanghai Institute of Microsystem and Information Technology, Chinese Academy of Sciences, Shanghai 200050, China}
\affiliation{State Key Laboratory of Surface Physics and Department of Physics, Fudan University, Shanghai 200433, China}

\author{Zheng Liu}
\affiliation{Department of Materials Science and Engineering, University of Utah, Salt Lake City, Utah 84112, USA}

\author{Yong-Shi Wu}
\affiliation{State Key Laboratory of Surface Physics and Department of Physics, Fudan University, Shanghai 200433, China}
\affiliation{Department of Physics and Astronomy, University of Utah, Salt Lake City, Utah 84112, USA}

\author{Yan Chen}
\affiliation{State Key Laboratory of Surface Physics and Department of Physics, Fudan University, Shanghai 200433, China}

\date{\today}


\begin{abstract}
Fractional Chern insulators (FCIs), having properties similar to those of the fractional quantum Hall effect, have been established numerically in various toy models. To fully explore their fundamental physics and to develop practical applications,  material realization is indispensable. Here we theoretically predict a realization of FCI in a two-dimensional organometallic material, which is known to have the prerequisite topological flat bands. Using numerical exact diagonalization we demonstrate that the presence of strong electronic correlations and fractional filling of such a system could lead to a rich phase diagram, including Abelian fractional quantum Hall (FQH), Fermi-liquid, and Wigner crystal states. In particular, the FQH state has been confirmed systematically by calculating the topological ground-state degeneracies, topological Chern number, and the quasihole excitation spectrum as well as the particle entanglement spectrum. Future experimental realization of the FQH state in such material may provide a route for developing practical applications of FCI with no need of the extreme conditions.
\end{abstract}

\maketitle

\section{Introduction}

Since the discovery of the fractional quantum Hall (FQH) effect~\cite{TsuiDC,RBLaughlin}, the emergence of exotic topological phases in strongly correlated systems has remained an object of intense interest both theoretically and experimentally. The FQH phase originates from the strong Coulomb interactions between electrons under a strong magnetic field. Recently, it has been proposed theoretically that the topological flat bands with nonzero Chern number could exist in the lattice model without external magnetic field ~\cite{ETang,KSun,TNeupert,XiangHu,FWang,Trescher}. These topological flat band models belong to the same topological class as the prototype Haldane model~\cite{Haldane} and are distinct from other flat bands with a zero Chern number~\cite{CWu}. A series of lattice models with nonzero Chern number and large energy gaps has been explicitly constructed, such as the kagome lattice model~\cite{ETang}, the checkerboard lattice model~\cite{KSun}, the honeycomb lattice model~\cite{TNeupert}, and the Ruby lattice model~\cite{XiangHu}. When such a topological flat band model with strong electronic correlations is partially filled (such as $\frac{1}{3}$ or $\frac{1}{5}$ filled), fractional Chern insulators (FCIs) can be realized with the appearance of the FQH phase~\cite{DNSheng,Bernevig,Venderbos1,XLQi,XLQi2,Venderbos2,YFWang,DXiao,Bernevig2,YFWang2,YLWu,TLiu,Zhaoliu,NYYao,Parameswaran}. However, all those studies are based on the toy models and it remains unclear whether such a FCI could exist in a real material.

From both theoretical and experimental points of view, to realize the FQH states in real materials three criteria must be required: (i) The band dispersion must be  quenched so that a nearly flat band has a large density of states at the Fermi level. (ii) The flat band structure should possess a nonzero Chern number, reflecting the underlying Berry phase accumulated by a particle moving in the band structure. (iii) One needs a strong Coulomb interaction which dominates over kinetic energies of the electrons. Because of these stringent criteria, no material to date has been experimentally observed to realize such a FCI.

Very recently, based on the LDA calculations a first-principles design of a two-dimensional (2D) indium-phenylene organometallic framework [see Fig.~\ref{fig:fig1}(a)] has been proposed to realize a nearly flat band state with a nonzero Chern number at around the Fermi level by combining lattice geometry, spin-orbit coupling, and ferromagnetism~\cite{ZLiu}. The key feature of this material structure is its similarity to  graphene, binding the $p$-orbital heavy elements (In) with organic ligands (parapheneylenes) into a hexagonal lattice. As a result, these states favor localization on the parapheneylenes ensuring a nearly flat energy band structure with spontaneous time-reversal symmetry breaking. Nevertheless, due to the In atoms having a large spin-orbit coupling interaction and considering the spontaneous time-reversal symmetry breaking, a particle moving in these bands accumulates a Berry phase that produces a nonzero Chern number. In addition, although the energy scales of both kinetic energy and the Coulomb interaction in organic materials are rather small, the ratio between the strength of Coulomb interaction and the kinetic energy is quite large. Thus, the above-mentioned three stringent criteria to establish a FCI might be satisfied in this material.

In this paper, we report a possible realization of exotic fractional topological states in a 2D organometallic material based on the exact numerical calculations for finite-size systems. The phase diagram is constructed, showing a rich structure with three phases that include states of an Abelian FQH, Fermi-liquid, and Wigner crystal. The existence of an Abelian FQH state is confirmed systematically using four independent techniques, such as the topological ground-state degeneracies, topological Chern number, and the quasihole excitation spectrum as well as the particle entanglement spectrum (PES). In particular, the PES can straightforwardly distinguish the three aforementioned phases according to different counting rules. At last, we also present a discussion of the possible experimental realization of an Abelian FQH state in such a 2D organometallic material.

The rest of this paper is organized as follows. In Sec.~\ref{two}, we present the theoretical model and method. The main theoretical results are discussed in Sec.~\ref{three}. Furthermore, the experimental realization is also discussed in Sec.~\ref{four}. Finally, in Sec.~\ref{five}, we summarize our main conclusions.

\section{Model and Method}
\label{two}

\begin{figure}[tbp]
\includegraphics[width=9cm,height=3.6cm]{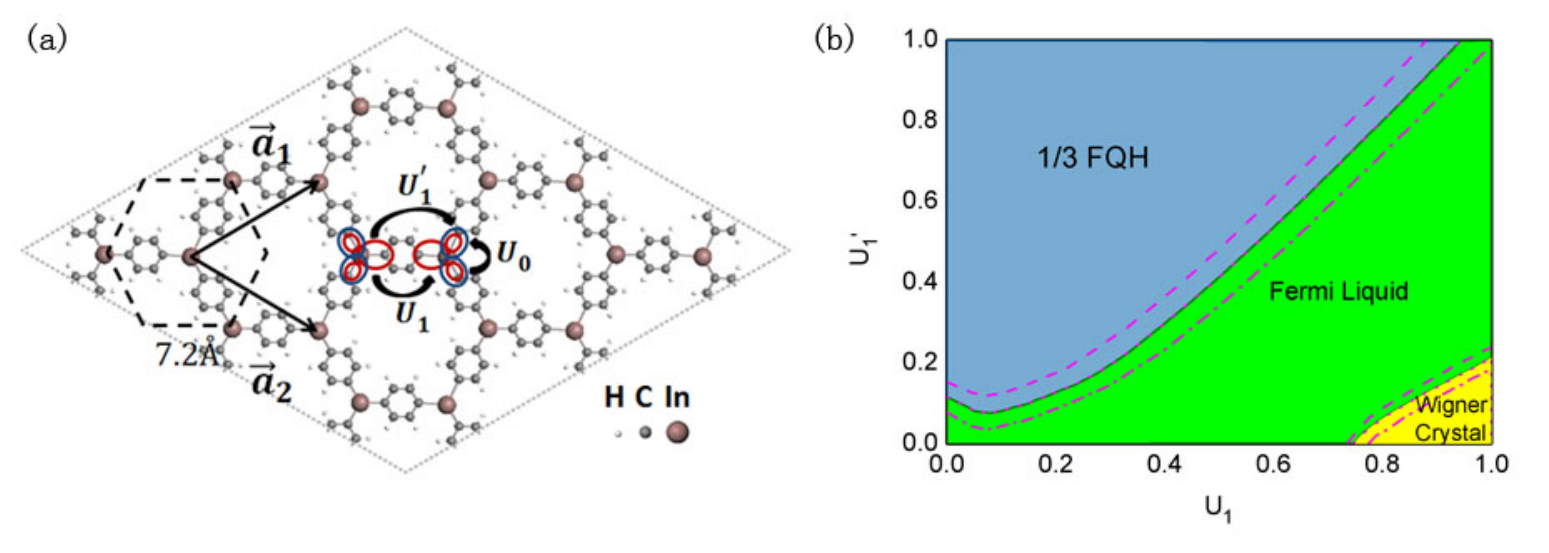}
\caption{(Color online). (a) The atomic structure of 2D indium-phenylene organometallic material. The red/blue contours represents the two molecular orbitals around an In atom, which are used as the basis to construct the effective Hamiltonian. $\vec{a}_{1,2}$ is the lattice vector. (b) The phases correspond to a FQH state (upper-left region), a Wigner crystal state (lower-right region), and a Fermi-liquid state (middle region) in the $U_1$-$U_1'$ plane at $\frac{1}{3}$-filling with the system size $N_s=48$. The dashed, dotted, and dashed-dotted lines represent the interaction $U_0$ with the values of 0.3, 0.4, and 0.5, respectively.}\label{fig:fig1}
\end{figure}

We consider the following Hamiltonian for a honeycomb lattice shown in Fig.~\ref{fig:fig1}(a):

\begin{eqnarray}
{\hat H} &=& {\hat H_{0}} + U_0\sum_{i,\alpha}\hat{n}_{i,\alpha}\hat{n}_{i,\bar{\alpha}} + U_1\sum_{\langle i,j \rangle,\alpha}\hat{n}_{i,\alpha}\hat{n}_{j,\alpha}
\notag\\
&+& U'_1\sum_{\langle i,j \rangle,\alpha}\hat{n}_{i,\alpha}\hat{n}_{j,\bar{\alpha}},
\label{eq:one}
\end{eqnarray}
where ${\hat H_{0}}$ describes an effective two-orbital tight-binding model for the honeycomb lattice fitting to the first-principles band structure defined in Ref.~\cite{ZLiu}:
\begin{eqnarray}\label{H0}
{\hat H_{0}}=-t_1\left(
  \begin{array}{cccc}
    0 & 0 & V_{xx} & V_{xy} \\
    0 & 0 & V_{xy} & V_{yy} \\
    V^*_{xx} & V^*_{xy} & 0 & 0 \\
    V^*_{xy} & V^*_{yy} & 0 & 0 \\
  \end{array}
\right)
+\lambda \left(
                     \begin{array}{cccc}
                       0 &-i & 0 & 0 \\
                       i & 0 & 0 & 0 \\
                       0 & 0 & 0 &-i \\
                       0 & 0 & i & 0 \\
                     \end{array}
                   \right),
\end{eqnarray}
in which $V_{xx}=\frac{1}{2}(1+e^{i\vec{k}\cdot \vec{a}_1})$, $V_{xy}=\frac{\sqrt{3}}{6}(1-e^{i\vec{k}\cdot\vec{a}_1})$, and $V_{yy}=\frac{1}{\sqrt{6}}(1+e^{i\vec{k}\cdot\vec{a}_1}+e^{i\vec{k}\cdot\vec{a}_2})$. $t_1$ describes the hopping parameter with strength $0.7$ eV and $\lambda$ denotes the spin-orbit coupling strength $0.05$ eV; $\vec{a}_{1,2}$ is the lattice vector as shown in Fig.~\ref{fig:fig1}(a).
$\hat{n}_{i,\alpha}$ is the on-site fermion particle number operator with orbital index $\alpha$ at site $i$. $U_0$ denotes the on-site Hubbard-like inter-orbital interaction, $U_1$ and
$U'_1$ corresponds to the nearest-neighbor intra- and inter-orbital interaction, respectively.

The honeycomb lattice has a unit cell with two sites and, thus, has four single-particle bands [see Eq. (\ref{H0})], where we have considered the spontaneous time-reversal symmetry breaking. 
The top and bottom bands in this system have unit Chern number with opposite sign whereas the middle two bands have zero Chern number. Indeed, any mixing between only two of lowest bands states would not change the Chern number of the lowest band, despite the gap between the two lowest bands being small~\cite{ZLiu}. If the interaction strength $U^{(')}_{0(1)}$ $\gg$ W (the band width of the lowest band $\approx 60$ meV), interaction effects dominate kinetic energy and partially filling the flat band would favor a strongly correlated state, such as a Wigner crystal or a FQH state. 
For our numerical study, we exactly diagonalize the many-body Hamiltonian given in Eq. (\ref{eq:one}) projected to the lowest flat band for a finite system with $N_1 \times N_2$ unit cell (total number of sites $N_s = 2 \times N_1 \times N_2$) with basis vectors shown in Fig.~\ref{fig:fig1}(a). We denote the number fermions as $N_e$, and define the filling factor as $\nu = \frac{N_e}{N_1N_2}$. With the periodic boundary condition implementing translational symmetries, we diagonalize the system Hamiltonian in each momentum $\mathbf{q}=(2\pi k_1/N_1, 2\pi k_2/N_2)$ sector with $(k_1, k_2)$ the associated integer quantum numbers.

\section{THEORETICAL RESULTS}
\label{three}

The effects of $U_1$ and $U_1'$ with fixed different values of the interaction $U_0$ are illustrated in a phase diagram, as shown in Fig.~\ref{fig:fig1}(b). At $\frac{1}{3}$ filling, the FQH state is most stable when $U'_1$ dominates. This is because the interaction $U'_1$ drives the system to be a spontaneous orbital symmetry breaking state resulting in all particles favor to occupy the same orbital, which is similar to that of the spin fully polarized Laughlin state~\cite{RBLaughlin}. Interestingly, at $\frac{1}{5}$-filling factor, similar results (not shown here) can also be obtained for five degenerate states if one includes the next nearest-neighbor repulsion~\cite{DNSheng}. Due to the competition in the orbital degrees of freedom between the interaction $U_1$ and $U_1'$, where interaction $U_1$ favors to drive the system to be a charge ordering state, the FQH state becomes less stable and a Fermi-liquid state may emerge by tuning the interaction strength $U_1$. If $U_1$ is further increased, the Fermi-liquid state is suppressed and a Wigner crystal state can appear. In addition, if we tune up the interaction strength $U_0$, all these phase boundaries move towards a larger value for $U_1$. Note that these phase boundaries are determined by evaluating the gap closing points of PES. 
On the other hand, we also performed the static structure factor [$n(\mathbf{k})$] calculations and shown that the Wigner crystal state is marked by sharp features in $n(\mathbf{k})$ at certain wave vectors, while liquid states including the FQH state and the Fermi-liquid state should be featureless in comparison to Wigner crystal state. In the following, all three phases will be carefully elaborated.

\subsection{FQH state}

\begin{figure}[tbp]
\includegraphics[bb=5 25 750 530, width=8.5cm, height=6.2cm]{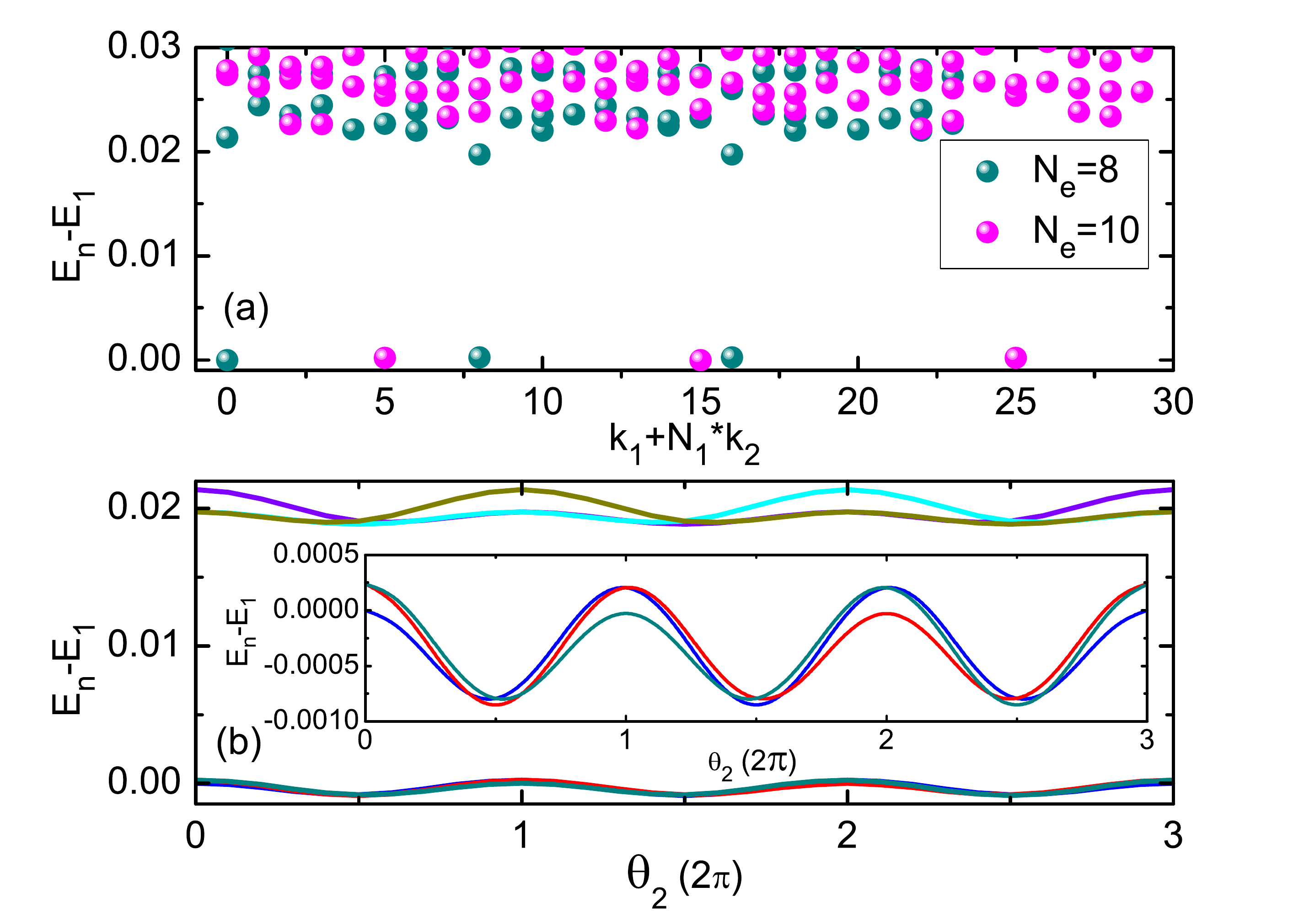}
\caption{(Color online). (a) Low-energy spectrum for $N_e = 8$ and 10, $N_1=N_e/2$, $N_2=6$. We only show the lowest excited level in each momentum sector in addition to the threefold ground state. (b) Evolution of the threefold degenerate ground state upon flux insertion along the $\hat{2}$-direction at $N_e=8$, $N_1=4$, and $N_2 = 6$. The threefold degenerate ground states flow into each other separated at each point in the flux insertion from the first excited state. Additionally, these energies are all shifted by $E_1$, which is the lowest energy for each system size. }\label{fig:fig2}
\end{figure}

In Fig.~\ref{fig:fig2}(a), the ground-state manifold is defined as the set of lowest states that are well separated from other excited states by a clear energy gap. Recall that this is a necessary condition for the $\nu=\frac{1}{3}$ fermionic FCI state. The energy gap is always significantly larger than the ground state splitting for various system sizes [see Fig.~\ref{fig:fig2}(a)]. Although for a finite system these states are not exactly degenerate, their energy difference should fall off exponentially as the system size increases. In addition, it is interesting to point out that if $(k_1, k_2)$ corresponds to the momentum sector for one of the states in the ground-state manifold, we find that the other state should be obtained in the sector $(k_1 + N_e, k_2 + N_e)$ [modulo $(N_1, N_2)$]. This correlation in the quantum numbers of the ground-state manifold implies that the Abelian FQH state has a topological nontrivial characteristic~\cite{DNSheng}.

To demonstrate that the ground states are indeed topologically nontrivial, we calculate the spectral flow under twisted boundary conditions, which amounts to inserting a magnetic flux through the cycles of the system. For a many-body state~\cite{QNiu}, $|\Phi(\mathbf{r}_j)\rangle$, the twisted boundary condition in the $\hat{1}(\hat{2})$ direction is $|\Phi(\mathbf{r}_j+N_{1(2)}\mathbf{a}_{1(2)})\rangle=e^{i\theta_{1(2)}}|\Phi(\mathbf{r}_j)\rangle$, where $\theta_{1(2)}$ is the boundary phase along the $\hat{1}$($\hat{2}$) direction and $\mathbf{a}_{1(2)}$ is the lattice vector. According to Laughlin's gauge argument~\cite{Laughlin,Halperin}, for the $\frac{1}{3}$-filling FQH system, when one adiabatically inserts  three quanta of flux, the states should evolve back to themselves looking exactly the same as before. In Fig.~\ref{fig:fig2}(b), considering the FQH ground-state manifold for the $\frac{1}{3}$-filling factor with $N_s = 48$, the three states are found to evolve into each other with level crossing and are separated from the other low-energy excitation by tuning the boundary phases. Eventually, energy levels evolve back to their initial configuration already after insertion of three flux quanta only, thus all three states share a total Chern number $\mathbf{1}$. The behavior of such spectral flows indicates that the quantized Hall conductance is $\sigma_H=\frac{1}{3}\frac{e^2}{h}$~\cite{TaoWu,QNiu}, which we have also confirmed by calculating the many-body Chern number.

\begin{figure}[tbp]
\includegraphics[bb=5 25 750 530, width=8.5cm, height=6.2cm]{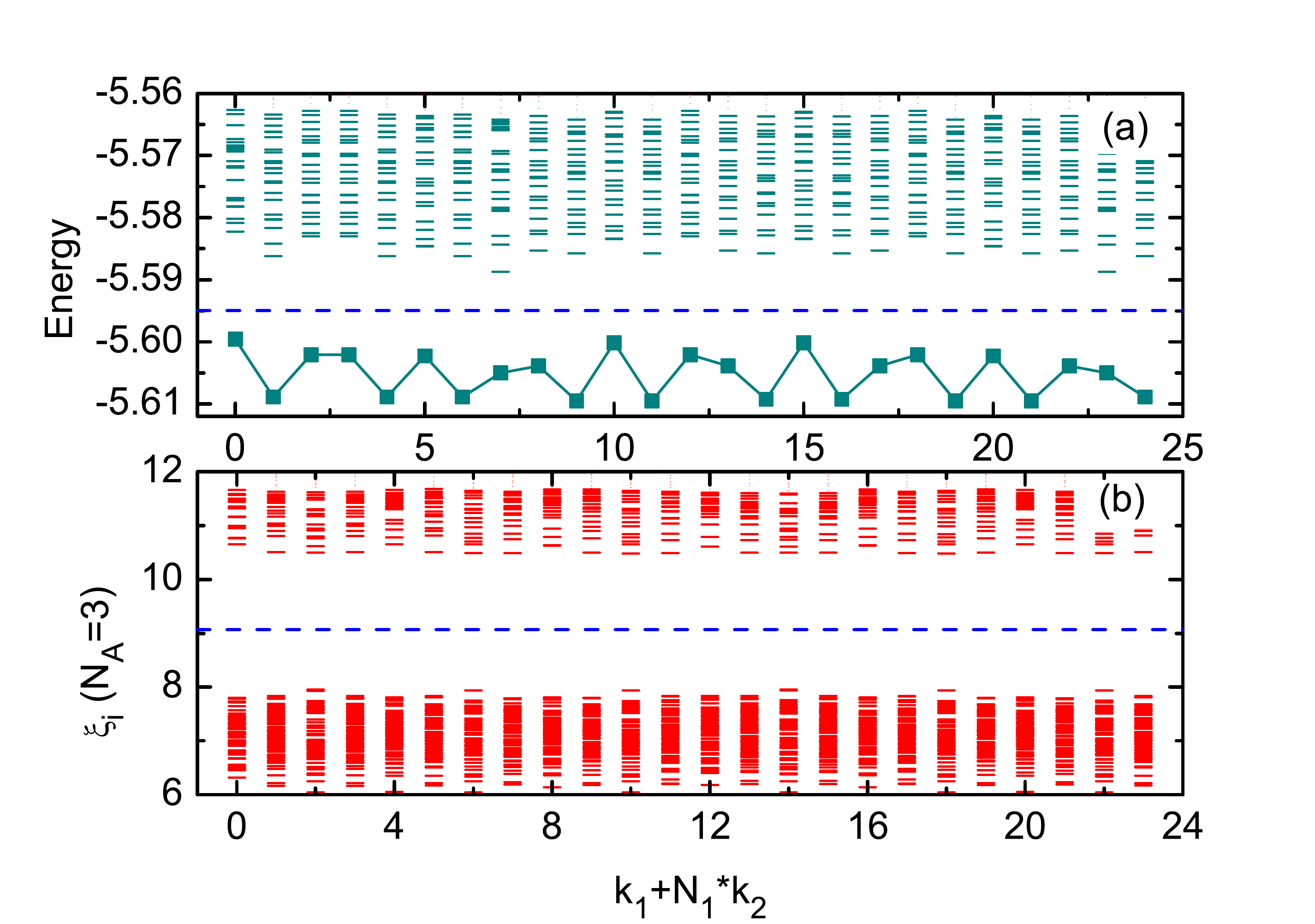}
\caption{(Color online). (a) Low-energy spectrum of the 2D organometallic material with $N_e = 8$ particles on the $N_s=50$ lattices. The number of states below the blue dashed line is 1 in all momentum sectors (one hole is added, 25 states below the gap), in agreement with the (1,3)-admissible counting rule. (b) PES probing the $N_A=3$ quasihole excitations of the $N_e = 8$ particles on the $N_s=48$ lattices. The total number of states below the blue dashed line is 1088, in agreement with the (1,3)-admissible counting rule.}\label{fig:fig3}
\end{figure}

To verify a possible fractional exclusion statistics~\cite{Haldane1991,Wu1994}, we turn to focus attention on the quasihole excitations, which are one of the most important characteristics for the FQH state in Landau levels. By keeping $N_e$ fixed and varying $N_1$ and/or $N_2$, we can add one hole into the system, as shown in Fig.~\ref{fig:fig3}(a). An energy gap is clearly visible in the quasihole excitation spectrum, and the total number of states below the gap has the same counting as predicted by the (1, 3)-admissible rule based on the generalized Pauli principle~\cite{Bernevig,FCZhang,SuWuYang}:
\begin{equation}
N_{FQH}^{N_{e}} = N_1N_2\frac{(N_1N_2 - 2N_{e} - 1)!}{N_{e}!(N_1N_2 - 3N_{e})!}.
\label{eq:eq2}
\end{equation}
For example, from the energy spectrum of the system with a single electron removed ($N_s=50$ and $N_e=8$), as shown in Fig.~\ref{fig:fig3}(a), one obtains $N_{FQH}^{N_{e}}=25$ from the analytic counting in Eq. (\ref{eq:eq2}), which agrees precisely with the number of states below the spectral gap. This further substantiates the claim that the ground state obtained at filling-$\frac{1}{3}$ indeed has the basic features of the Laughlin's FQH state.

To further corroborate our finding regarding the FQH state, we have also investigated the PES~\cite{Bernevig,Bernevig2,YLWu,TLiu,ZLiu,Sterdyniak,Chandran}, which provides an independent
signature of the excitation structure of the system and can be implemented to rule out other possibilities, such as a Wigner crystal state~\cite{Bernevig_CDW}. Using this powerful tool, we provide further evidence to differentiate the nature of the ground state at the $\frac{1}{3}$-filling, whether this is a Laughlin FQH state or a Wigner crystal state. We note here that concrete expressions for the model wave functions have not yet been established for FCIs. Thus this PES tool becomes highly valuable in this instance because no overlap integrals with model wave functions can be computed. Specifically, we partition the system in the way described in Ref~\cite{Bernevig} and divide the $N_e$ particles into two subsystems
of $N_A$ and $N_B$ particles, and trace out the degrees of freedom carried by the $N_B$ particles. The eigenvalues $e^{-\xi}$ of the resulting reduced density matrix $\rho_A=$Tr$_B\rho=$Tr$_B|\Phi\rangle\langle\Phi|$. The entanglement energy levels $\xi$ can then be displayed in groups labeled by the total momentum $(k_1,k_2)$ for
the $N_A$ particles.
As shown in Fig.~\ref{fig:fig3}(b), the spectrum is very similar to that found in previously studied FCIs~\cite{Bernevig,Bernevig2,YLWu,ZLiu}. We observe a clear entanglement gap in the spectrum separating these levels from generic ones and the counting of the entanglement energy levels below the gap matches the (1,3)-admissible quasihole counting of $N_A$ particles of FCIs on the $N_1\times N_2$ reciprocal lattice [see Eq. (\ref{eq:eq2})].

\begin{figure}[tbp]
\includegraphics[bb=40 25 750 530, width=7.5cm, height=5.5cm]{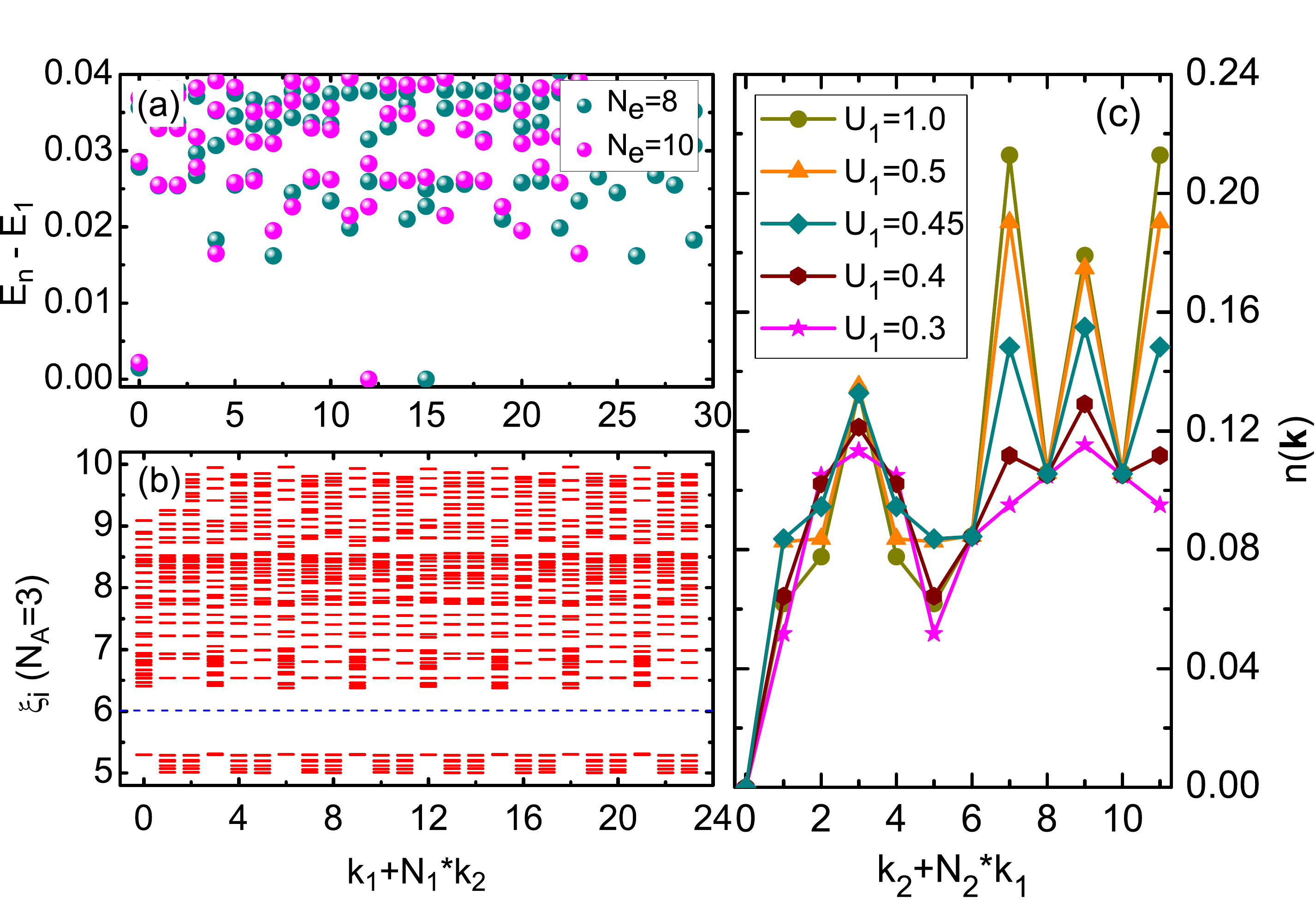}
\caption{(Color online). (a) Low-energy spectrum for $N_e = 8$ and 10, $N_1=3$, $N_2=N_e$. We only show the lowest excited level in each momentum sector in addition to the threefold ground state. (b) PES probing the $N_A=3$ quasihole excitations of the $N_e = 8$ particles on the $N_s=48$ lattices. The total number of states below the blue dashed line is 112. (c) Static structure factor $n(\mathbf{k})$ as a function of interaction $U_1$ with fixing $U_0$=$U'_1$=0 and the system size $N_s$=24, where the two sharp peaks indicate the formation of a CDW state.}\label{fig:fig4}
\end{figure}

\subsection{Wigner crystal state}

In Fig.~\ref{fig:fig4}(a), the ground-state manifold is twofold and well separated from other excited states by a clear energy gap, similar to that for the FQH state. However, when tuning the boundary phases using the twist boundary condition, the states have not been found to evolve into each other with level crossing, and levels return to their initial configuration just after insertion of a single flux quantum. This behavior is very different from that for the FQH state. To distinguish between FQH state and this one, we further calculate the PES [see Fig.~\ref{fig:fig4}(b)]. The number of states below the entanglement gap in the PES is much smaller than the FQH counting [see Eq. (\ref{eq:eq2})], but matches the charge density wave (CDW) counting~\cite{Bernevig_CDW,Budich,Alexander}:$N_{CDW}^{N_A} = 2\left(\begin{array}{c}
                   N_e \\
                   N_A
                 \end{array}\right)=2\left(\begin{array}{c}
                   8 \\
                   3
                 \end{array}\right)=112$ [see Fig.~\ref{fig:fig4}(b)].
Therefore, we conclude that this state indeed is a Wigner crystal state.

On the other hand, we also calculate the key properties of the corresponding ground states in real space without projection onto the single-particle bands, such as the static (charge-) structure factor~\cite{SSF1,SSF2}, defined as:
\begin{eqnarray}
n(\mathbf{k})&=&\frac{1}{N_s}\sum_{\alpha,\beta}\sum_{j,l}^{N_s}e^{i\mathbf{k}\cdot(\mathbf{R}_{j}-\mathbf{R}_l)}(\langle \Phi| \hat{n}_{j,\alpha} \hat{n}_{l,\beta}|\Phi \rangle
 \notag\\
&& - \langle \Phi|\hat{n}_{j,\alpha} |\Phi\rangle\langle\Phi|\hat{n}_{l,\beta} |\Phi\rangle).
\label{eq:eqSOM}
\end{eqnarray}
Charge-density modulations are marked by sharp features in $n(\mathbf{k})$ at certain wave vectors, as shown in Fig.~\ref{fig:fig4}(c). It is evidenced that there is a typical feature of Wigner crystal state~\cite{SSF1,SSF2}: The two peaks appear in $n(\mathbf{k})$ at momenta $\mathbf{k}=\pm K$ and increase continuously upon increasing interaction $U_1$, which further justifies that this state is indeed a Wigner crystal state. 


\subsection{Fermi-liquid state}

For the Fermi-liquid phase [see Fig.~\ref{fig:fig1}(b)], the ground-state calculations evidenced that there is no well-defined nearly degenerate ground-state manifold or spectrum gap and the obtained total Chern number varies with interaction parameters (for example, $U_0$, $U_1$, and $U'_1$), suggesting the absence of topological properties~\cite{DNSheng}. The PES shows that there is no counting rule with invisible entanglement gap (the gap of entanglement spectrum closed). Furthermore, the calculated static structure factor also evidences that the $n(\mathbf{k})$ remains almost unchanged upon variation of interaction parameters. All those properties are the features of the Fermi-liquid state.

\section{Experimental realization and detection}
\label{four}

Further investigation beyond the standard density-functional formalism is required to determine the exact values of the microscopic parameters $U_0$, $U_1$, $U_1'$, as well as the ground state of the 2D organometallic material. We shall address this issue in future studies. Nevertheless, our phase diagram provides a general guidance to produce the sought FQH state. One intriguing implication drawn from the phase diagram [see Fig.~\ref{fig:fig1}(b)] is that the FQH state appears in the upper-left region of the $U_1$-$U_1'$ parameter space, which corresponds to a larger inter-orbital interaction $U_1'$ than the intra-orbital interaction $U_1$. General speaking, the strengths of interactions ($U_0$, $U_1$, $U_1'$) as well as bare band dispersions are determined by the orbital wavefunctions and the (nonlocal) dielectric function. Owing to the remarkable flexibility of the organic systems, these microscopic parameters can be tuned by various kinds of chemical ways, e.g. by functionalizing the benzenes with different chemical groups, replacing the benzenes with other organic ligands, or using different metal atoms. By the application of hydrostatic pressure or uniaxial stress, these microscopic parameters can be tuned as well. We expect that both the Abelian FQH and Wigner crystal states can be realized in such materials under different chemical and/or physical conditions.
Since the existence of FCI state in 2D organometallic material does not require an external magnetic field and may potentially persist to quite high temperatures,
it becomes easier to detect and then characterize the Abelian FQH state as well as Wigner crystal state experimentally in such material.

\section{Conclusion}
\label{five}

We have performed the exact diagonalization on the study of the electronic many-body effects on the nearly flat band structure with nonzero Chern number in 2D organometallic material. The phase diagram is constructed and shows that this system exhibits rich phases, including an Abelian fractional quantum Hall state, a Wigner crystal state and a Fermi-liquid one, by tuning the interactions strength. These exotic states can be distinguished by calculating the PES according to different counting rules. At last, we also discuss the possible experimental realization of an Abelian FQH state in such 2D organometallic material.

\section*{Acknowledgement}

The authors thank R. B. Tao, D. N. Sheng, T. K. Lee, S. Q. Shen, A. Tsevlik, X. M. Xie, and Z. C. Gu for invaluable discussions. This work was supported by the Strategic Priority Research Program (B) of the Chinese Academy of Sciences (Grant No. XDB04010600) and the National Natural Science Foundation of China (Grant No. 11227902) (W.L.), by the State Key Programs of China (Grant Nos. 2012CB921604 and 2009CB929204)
and the National Natural Science Foundation of China (Grant Nos. 11074043 and 11274069) (Y.C.), by the the DOE-BES (DE-FG02-03ER46027) (Z.L.), and by the U.S. NSF Grant No. PHY-1068558 (Y.S.W.).

\end{document}